\RequirePackage{fix-cm}

\documentclass[twocolumn]{svjour3}          
\usepackage{amsmath,amssymb}
\smartqed  
\usepackage{url}
\usepackage[%
  colorlinks=true,
  urlcolor=blue,
  linkcolor=blue,
  citecolor=blue
]{hyperref}
\hypersetup{linkcolor=red, citecolor=red, colorlinks=true}

\usepackage{mathtools,amsmath}

\begin{document}

\title{Diffusion plays an unusual role in ecological  quasi-neutral competition in metapopulations}

\author{Marcelo A. Pires         \and
        Nuno Crokidakis \and Silvio M. Duarte Queir\'os 
}


\institute{Marcelo A. Pires \at
              Centro Brasileiro de Pesquisas F\'isicas, Rio de Janeiro/RJ, Brazil \\
              \email{piresma@cbpf.br}           
           \and
           Nuno Crokidakis \at
              Instituto de F\'isica, Universidade Federal Fluminense, Niter\'oi/RJ, Brazil \\
              \email{nuno@mail.if.uff.br}
           \and
           Silvio M. Duarte Queir\'os \at
              Centro Brasileiro de Pesquisas F\'isicas and Instituto Nacional de Ci\^encia e Tecnologia de Sistemas Complexos INCT-SC, Rio de Janeiro/RJ, Brazil   \\
              \email{sdqueiro@cbpf.br}
}

\date{Received: date / Accepted: date}

\maketitle

\begin{abstract}

We investigate the phenomenology emerging from a 2-species dynamics under the scenario of a quasi-neutral competition within a metapopulation framework. 
We employ stochastic and deterministic approaches, namely spatially-constrained individual-based Monte Carlo simulations and  coupled mean-field ODEs. Our results show the  multifold interplay between competition, birth-death dynamics and spatial constraints induces a nonmonotonic relation between the  ecological majority-minority switching and the diffusion between patches. This means that diffusion can set off birth-death ratios and enhance the preservation of a species.

\end{abstract}

\section{\label{sec:intro}Introduction}

The battle for resources plays a significant role in the dynamics of competitive ecosystems. For a long time, the outcome of such a dispute was directly associated with the set of birth/death ratios of the contending species, $\lambda _{i}/\alpha_{i}$. However, that scenario has been challenged by ecological models seasoned with other factors such as mobility, which proved themselves capable of leading to a priori upset results. To a large extent, those counter-intuitive\footnote{When one only takes $\lambda _{i}/\alpha_{i}$ into consideration.} results come to pass due to the competition established between the different factors leading to the emergence of non-linear effects or even simple correlations --- as described by the Allee effect
\footnote{The Allee effect can be defined as the positive correlation between the absolute average individual fitness in a population and its size over some finite interval. It can be separated into its strong and weak versions. The former corresponds to the case when the deviation from the logistic growth includes an initial population threshold below which the population goes extinct whereas the latter treats positive relations between the overall individual fitness in the population density and does not present population size nor density thresholds.}
--- that overcome the initial reasoning on which Gause's law is based. The impact of those different contributions to competitive dynamics is especially interesting when one is dealing with quasi-neutral instances, for which the specific values of the birth, $\lambda _{i}$, and death, $\alpha_{i}$, rate of species $i$ yields the same ratio $\lambda _{i}/\alpha_{i}$ for all $i$. As we explore herein later on, besides the standard deterministic approach to an ecosystem, the problem has been analyzed from a stochastic perspective by means of a series of techniques systematically applied at the population scale.

In the present work, we tackle the problem of understanding the role played by patch diffusion --- which we use as a quantitative proxy for mobility --- in quasi-neutral competition within the metapopulation framework. Ecologically, a metapopulation --- i.e., a population of populations --- corresponds to a group of local connected populations of a species, the size of which changes in time due to microscopic factors such as the birth, death and migration of the individuals as well as mesoscopic events affecting the local populations contained within the metapopulation, namely emergence and dissolution. We have considered a survey at this scale because a small local population can imperil the species (e.g., by reducing mating)~\cite{thompson2016}. Besides being empirically observed~\cite{sweanor2001meta,borthagaray2015meta,fobert2019meta}, metapopulation approaches have set forth important results regarding ecological landscape dynamics in either homogeneous~\cite{johst2002meta,vuilleumier2006meta,colombo2015meta} or heterogeneous populations \cite{nagatani2019meta,souza2014meta,juher2009meta}. Our results show that the interplay between quasi-neutral competition between two species with different biological clocks, spatial constraints and diffusion in metapopulation is complex. Indeed, we verified that large mobility between different patches can have the same impact as no migration between patches. In addition, depending on the level of mobility, being biologically slower can be actually an advantage. The ecological majority-minority switching exhibits a nonmonotonic relation with the diffusion between patches.

The remaining of this manuscript is organized  as follows: as previously mentioned, in Sec.~\ref{sec:lit-rev}, we present a list of correlated works; in Sec.~\ref{sec:model}, we introduce our model as well as its algorithmic implementation; in Sec.~\ref{sec:results}, we deliver our results obtained by both analytical and computational approaches; and in Sec.~\ref{sec:fin-rem}, we address our conclusions and remarks as well as perspectives for future work.

\section{\label{sec:lit-rev}Literature review}

The effects of diffusion were studied in many works in recent years: Smith \textit{et. al.} ~\cite{smith2014programmed} studied the Allee effect in bacteria populations and showed that it led to a biphasic dependence of bacterial spread on the dispersal rate: spread is promoted for intermediate dispersal rates but inhibited at low or high dispersal rates. Correlated to such experimental work, the authors in ~\cite{pires2019optimal} explored theoretically the threefold interplay among the Allee Effect, dispersal, and spatial constraints. They showed that the  survival-extinction boundary undergoes a novel transition of monotonicity  in the way that for the  nonmonotonic regime there is an optimal dispersal rate that maximizes the  survival probability. Diffusion of populations can also relate to the emergence of Parrondo's paradox instances for which the combination of two losing (extinction) strategies -- diffusion and inefficient $\alpha$ -- combined yield a winning (preservation) situation
~\cite{tan2017nomadic,tan2019parrondo,tan2020predator}.

Considering competition between two distinct species, Pigolotti and Benzi showed that an effective selective advantage emerges when the two competing species diffuse at different rates ~\cite{pigolotti2014selective}. In reaction/kinetic systems, diffusion can lead to distinct scenarios in reaction/kinetic system: it destroys the stability of possible equilibrium, leading to the formation of characteristic patterns; drive an otherwise persistent competing species to extinction ~\cite{su2019rich}. 
Some paradoxical situations can emerge in the competition between species as well. For instance, for a sizable range of asymmetries in the growth and competition rates, it was discussed that the numerically disadvantaged species according to the deterministic rate equations survive much longer ~\cite{gabel2013survival}.
In  $d$-dimensional spatial structures, the survival of the scarcer in space is verified for situations in which the more competitive species is closer to the threshold for extinction than is the less competitive species when considered in isolation ~\cite{dos2013survival}.

Another recent work studied the competition between fast- and slow-diffusing species, considering non-homogeneous environments ~\cite{pigolotti2016competition}. The authors considered the case in which non-homogeneity in the nutrients is contrasted with a fluid flow concentrating individuals around a velocity sink. In such a case, diffusing faster constitutes an advantage as faster individuals can colonize more easily upstream regions, from which they can invade. It was also argued that in time-independent environments it is always convenient to diffuse less
~\cite{pigolotti2016competition,hastings1983can,dockery1998evolution}; particularly, the authors in Ref. ~\cite{pigolotti2016competition} suggested that deterministic models can miss a crucial ingredient to determine the best dispersal strategy.

Considering two species that differ only in the rates of their biological clocks, the authors in \cite{de2017advantage} showed that the slower species can enjoy an advantage in stationary population density for reproduction rates close to (but greater than) a critical value, and large initial population densities. Alternatively, it was shown~\cite{cheong2018clock} that switching-rule approaches relying upon biological clocks provide an efficient mechanism by which species might undergo behavioral nomadic-colonial alternation that allows them to develop well. For a recent review about the \textit{slower is faster} effect, considering pedestrian dynamics, vehicle traffic, traffic light control, logistics, public transport, social dynamics, ecological systems, and others, see~\cite{gershenson2015slower}.

\section{\label{sec:model}Model}

Consider a metapopulation
with  $L$ subpopulations composed of agents that are able to move, die, and reproduce. 
As usual in metapopulation dynamics, 
we consider well-mixed subpopulations so that the individuals inside each of them can interact with one another, or in Statistical Physics parlance, we employ a local dynamics that has a mean-field character.  The mobility is implemented as a random walk between the neighbor subpopulations and it occurs with probability $D$ for each agent.

At each time step, if the diffusion event is not chosen with probability $1-D$, then we implement the events related to the quasi-neutral competition~\cite{de2017advantage} between species $F$ and $S$ inside each subpopulation:
\begin{align}
& F+V\Rightarrow 2F   \quad \text{ with rate } \lambda_F 
\\ 
& F\Rightarrow V  \quad \text{ with rate } \alpha_F
\\ 
& S+V\Rightarrow 2S   \quad \text{ with rate } \lambda_S
\\ 
& S\Rightarrow V  \quad \text{ with rate } \alpha_S,
\end{align} 
where $F$ ($S$) stands to faster (slower) species and V for a vacancy.

In order to allow a comparison between the biological clock of the two species we introduce a relative birth ratio so that
\begin{equation}
\lambda_S = r \lambda_F, \qquad
\alpha_S = r \alpha_F; 
\end{equation}
explicitly we have the following relations:
\begin{equation*}
\begin{cases} 
\lambda_S = r \lambda_F  &  r<1 \text{ disadvantage of S:  smaller  birth rate,}
\\ 
\alpha_S = r \alpha_F  &   r<1 \text{ advantage of S: smaller death rate.}
\end{cases}
\end{equation*}

At this point, we would like to emphasize that the aim of this work is not to model a specific biological dynamics, but rather to carry out a systematic investigation over the range of arising scenarios from this minimal agent-based competition-reproduction-death dynamics with dispersal,  which can be tailored with additional features that match the traits of a determined ecological system. This sort of approach employing a minimal model allows one to grasp the key mechanisms that will be present in more realistic models seasoned for particular systems.

To computationally implement the set of microscopic relations Eqs.~(1)-(4), we consider an array with $N$ states divided into the $L$ subpopulations. In this case, we assume periodic boundary conditions. 
Each state in subpopulation $u$ indicates an agent $\{i_F^u,i_S^u\}$  or a vacancy, $i_V^u$.
Our time unit is a Monte Carlo step (mcs) that consists of a visit to each one of the $N$ states.

\begin{figure*}[!hbtp]
\centering 
\includegraphics[width=0.99\linewidth]{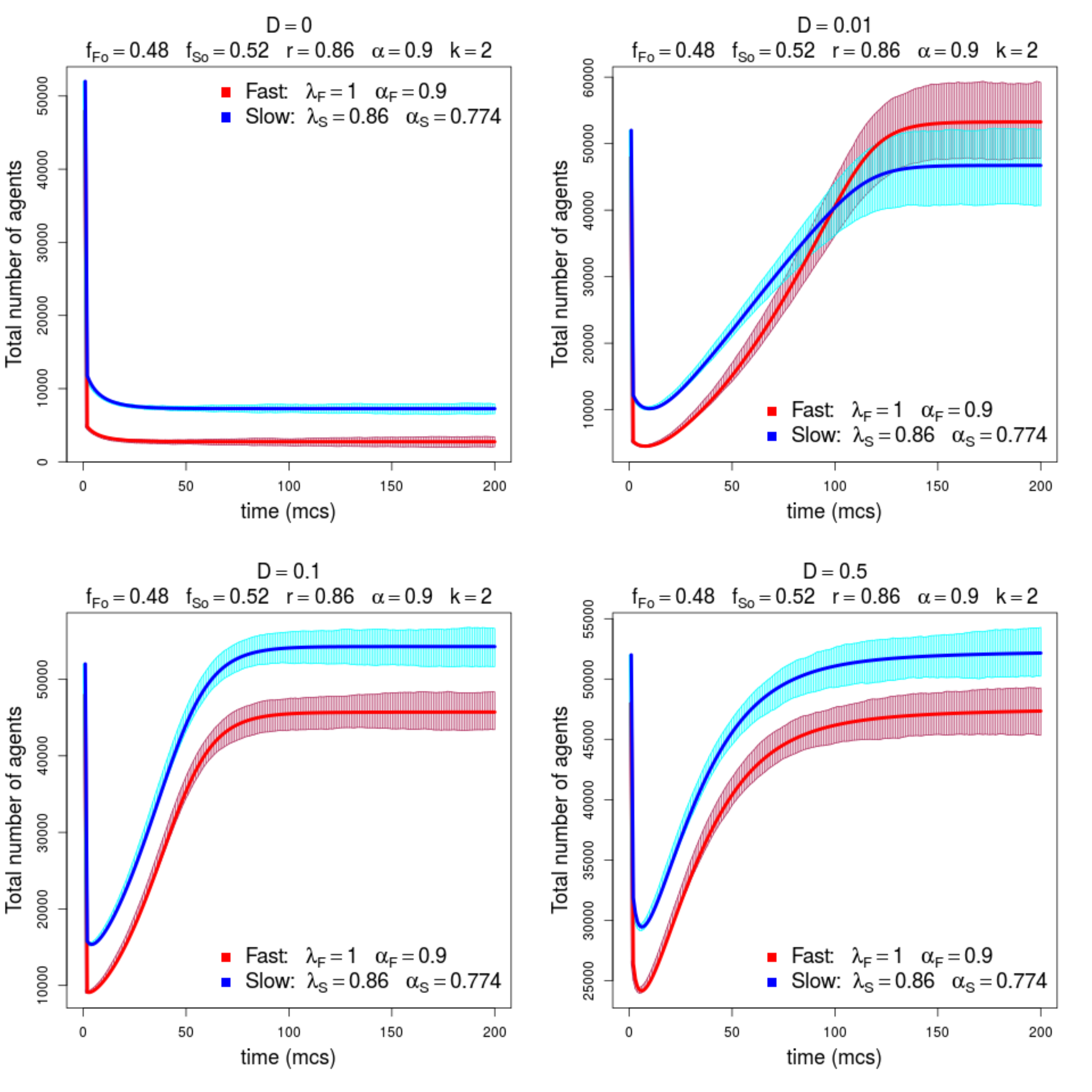}
\caption{(scenario I) Time series for the total number of individuals F and S. Shaded area comes from Monte Carlo simulation (mean$\pm$standard deviation). The theoretical lines  comes from the numerical solution of Eqs.~(\ref{Eq:Vu2})-(\ref{Eq:Bu2}). To summarize:  $D=0$, winner: S; $D=0.01$, winner: F; $D=0.1$, winner: S; $D=0.5$, winner: S. }
\label{Fig:alpha_vs_D_k_1}
\end{figure*}

\begin{figure*}[!hbtp]
\centering 
\includegraphics[width=0.99\linewidth]{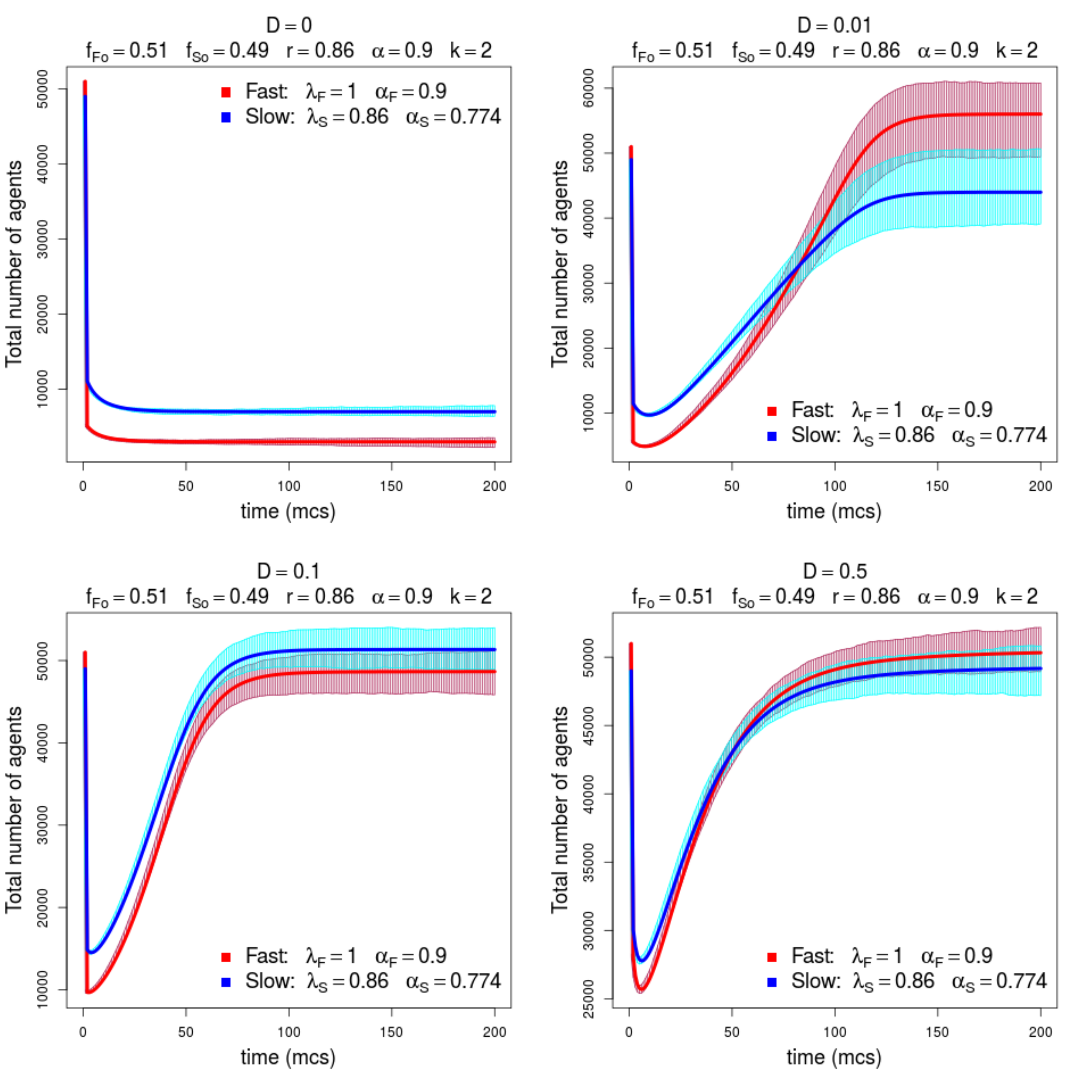}
\caption{(scenario II) Time series for the total number of individuals F and S. Shaded area comes from Monte Carlo simulation (mean$\pm$standard deviation). The theoretical lines  comes from Eqs.~(\ref{Eq:Vu2})-(\ref{Eq:Bu2}). To summarize: $D=0$, winner: S; $D=0.01$, winner: F; $D=0.1$, winner: S; $D=0.5$, winner: F. }
\label{Fig:alpha_vs_D_k_2}
\end{figure*}

For each state $i=1,\ldots,N $:
\begin{itemize}
\item First we get the subpopulation, say $u$, of the state $i$.

\vspace*{0.19cm}
\item With probability  $D$:
 \begin{itemize}

 \vspace*{0.19cm}
\item \textbf{Diffusion:}
\newline
$i_F^u \Rightarrow i_F^w $.  (event 1F)
\newline
$i_S^u \Rightarrow i_S^w $.  (event 1S)
  \end{itemize}

\vspace*{0.19cm}
\item and with probability $1-D$:
\begin{itemize}

\vspace*{0.19cm}
\item \textbf{Reproduction:}
\newline
rate $\lambda_F$:
 $i_V^u+j_F^u \Rightarrow   i_F^u+j_F^u $. (event 2F)
\newline
rate $\lambda_S$:
 $i_V^u+j_S^u \Rightarrow   i_S^u+j_S^u$. (event 2S)

\vspace*{0.19cm}
\item \textbf{Death:} 
\newline
rate $\alpha_F$:
 $i_F^u \Rightarrow i_V^u$. (event 3F)
\newline
rate $\alpha_S$:
 $i_S^u \Rightarrow i_V^u$. (event 3S)

  \end{itemize}
  
  \end{itemize}

The details of our computational approach can be found at the URL: \url{https://github.com/PiresMA/diffusion_2cp}.

\section{\label{sec:results}Results}


Applying  a previous  bottom-up  mathematical framework ~\cite{pires2019optimal} to the set of rules described above we arrive at
\begin{equation}
 \frac{dV_u}{dt}
 =
 (1-D)  
 \Big[
\overbrace{-\frac{\lambda V_u (F_u + r\, S_u) }{V_u+F_u+S_u}}^\text{Reproduction}
+
\overbrace{\alpha (F_u + r\, S_u)}^\text{\newline Death}         
\Big] 
\label{Eq:Vu2}             
\end{equation}

\begin{multline}
  \frac{dF_u}{dt}
 =
 (1-D)  
 \Big[
\underbrace{\frac{\lambda V_u F_u}{V_u+F_u+S_u}}_\text{Reproduction}
-
\underbrace{\alpha F_u}_\text{\newline Death}         
\Big] 
+ \\
D\Big[
\underbrace{-F_u}_\text{Emigration}
+
\underbrace{\sum_{z=1}^{L} \frac{1}{k} W_{uz} F_{z}}_\text{Immigration}
\Big]              
\label{Eq:Au2}
\end{multline}

\begin{multline}
  \frac{dS_u}{dt}
 =
 (1-D)  
 \Big[
\underbrace{\frac{r\lambda V_u S_u}{V_u+Fu+S_u}}_\text{Reproduction}
-
\underbrace{r\alpha S_u}_\text{\newline Death}         
\Big] 
\\
D\Big[
\underbrace{-S_u}_\text{Emigration}
+
\underbrace{\sum_{z=1}^{L} \frac{1}{k} W_{uz} S_{z}}_\text{Immigration}
\Big],              
\label{Eq:Bu2}
\end{multline}
where $W_{uz}$ is the adjacency matrix which assumes values 1 if $u$ and $z$ are connected or 0 otherwise. We work with a circular/ring metapopulation wherein
each location/region contains a population that is coupled to $k$ neighbor populations. The parameter $k$ is the  connectivity of each population,  i.e., it regulates the strength of the spatial constraints. 
We use an initial condition given by 
\begin{align}
F_u(0) &= 
\frac{f_{Fo}}{n_s} 
\frac{N}{L}  
\quad\quad u={1,\ldots,n_s}
\\
S_u(0) &= 
\frac{f_{So}}{n_s} 
\frac{N}{L}  
\quad\quad u={1,\ldots,n_s}
\end{align}
with $f_{Fo}, f_{So}$ being the modulating factors that account for the fraction of individuals at the subpopulations, $N/L$ is the initial size of each subpopulation and $n_s$ is the number of initial sources.  Additionally, we use $V_u(0)=N/L-F_u(0)-S_u(0)$.
As we have a plethora of parameters we set $f_{So}=1-f_{Fo}$ as well as $n_s=1$ (all the agents are located initially in one source). Besides, without losing generality we have also set $\lambda=1$. This is equivalent to reescalate all the variables with the reproduction rate $\lambda$.

We implement our spatially-constrained Monte Carlo algorithm in computer simulations considering 
metapopulations with $N=10^6$ and $L=10$.  Despite that fact, we assert that  all of the our findings 
remain valid for larger metapopulations
since the deterministic coupled mean-field Eqs.~(\ref{Eq:Vu2})-(\ref{Eq:Bu2}) are valid in the  limit of infinite population.  This assumption is clearly validated with the results shown in Figs.~\ref{Fig:alpha_vs_D_k_1}-\ref{Fig:alpha_vs_D_k_2} where we see a good agreement between  the  numerical solution of the multidimensional ODEs in  Eqs.~(\ref{Eq:Vu2})-(\ref{Eq:Bu2}) 
and the individual-based Monte Carlo simulations~\cite{vincenot2011theoretical,grimm2005individual} with 100 samples in each panel.

In Fig.~\ref{Fig:alpha_vs_D_k_1}-\ref{Fig:alpha_vs_D_k_2}, we present the outcomes for some specific configurations in order to explain in detail the myriad of majority-minority switching in the ecological dynamics.  In the  subsequent analyses we show the results for more general settings in order to provide an overall perspective about the robustness of the emergent phenomenology.

Focusing solely on Fig.~\ref{Fig:alpha_vs_D_k_1}, we depict the time  evolution of the number of individuals F and S for which the initial population of the faster species (initial density $f_{Fo}=0.48$) is smaller than the slower one (initial density $f_{So}=0.52$). 
For $D=0$, the slower species S becomes dominant at the steady state. For small values of diffusion (such as $D=0.01$) the initially majority species (S) becomes the minority one at the steady state. 
Such scenario changes for intermediate values of the diffusion rate ($D=0.1$), where we see that S recover the majority position.
For $D=0.5$, the dominance of species S becomes greater.
From the different panels, it is visible that, as we increase the value of the diffusion parameter $D$, we first observe the faster species achieves a larger final (steady state) population, and afterward, the slower species S becomes prevalent.

However, the outcome changes as we pay attention to Fig.~\ref{Fig:alpha_vs_D_k_2}, where the faster species has the initial majority ($f_{Fo}=0.51$ and $f_{So}=0.49$). 
For $D=0$, the slower species S becomes again the dominant one at the steady state. For  $D=0.01$ the initially prevailing species (F) becomes the minority one in the short run but recovers to become the majority at the steady state. 
For intermediate values in the mobility ($D=0.1$) the minority species S becomes the majority.
For $D=0.5$  the dominance of species S is destroyed again.
That is, the overall picture now is more diverse than the previous setting in Fig.~\ref{Fig:alpha_vs_D_k_1}.

\begin{figure}[!hbtp]
\centering 
\includegraphics[width=0.99\linewidth]{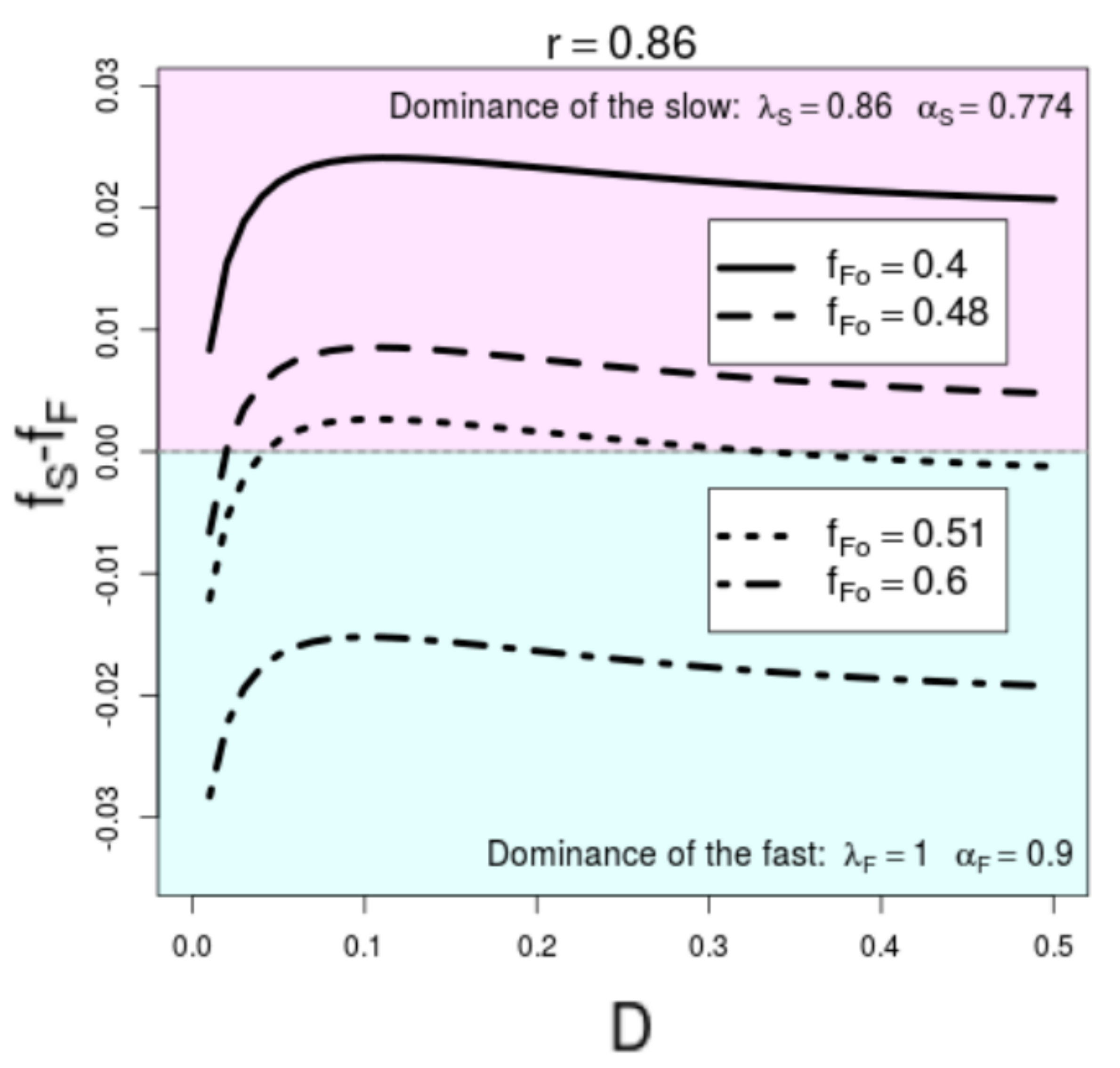}
\caption{Relative difference in the number of agents $f_S-f_F$ versus the diffusion parameter $0<D\leq 0.5$. Results are for $r=0.86$. Diffusion leads to four types of distinct scenarios regarding the dominance of the species F/S.}
\label{Fig:tot_vs_D_fao}
\end{figure}

The dominance of a species can be also represented by the
relative difference between the size of the populations at the steady state, $\Delta_f \equiv f_S-f_F =  (N_S-N_F)/N $. For a given value of the asymmetry parameter  $r$ we can plot parametric curves for fixed initial conditions  in the  \mbox{$\Delta_f$-$D$}
plane depicted in Fig.~\ref{Fig:tot_vs_D_fao}. In that plot, we verify that taking into consideration the initial condition regarding the initial fraction of each species, there are intuitive curves for which the majority species is the prevalent species in the final (steady) state (e.g., solid and dot-dashed lines) whereas for the dotted line we find a trivial region for large values of the diffusion parameter, $D\gtrsim 0.3$, as well as very limited diffusion parameter $D \lesssim 0.05$ and within those values of $D$ we observe a non-trivial region where in spite of being outnumbered at first by $F$, the slower species $S$ reach a larger population at last.

\begin{figure}[!hbtp]
\centering 
\includegraphics[width=0.99\linewidth]{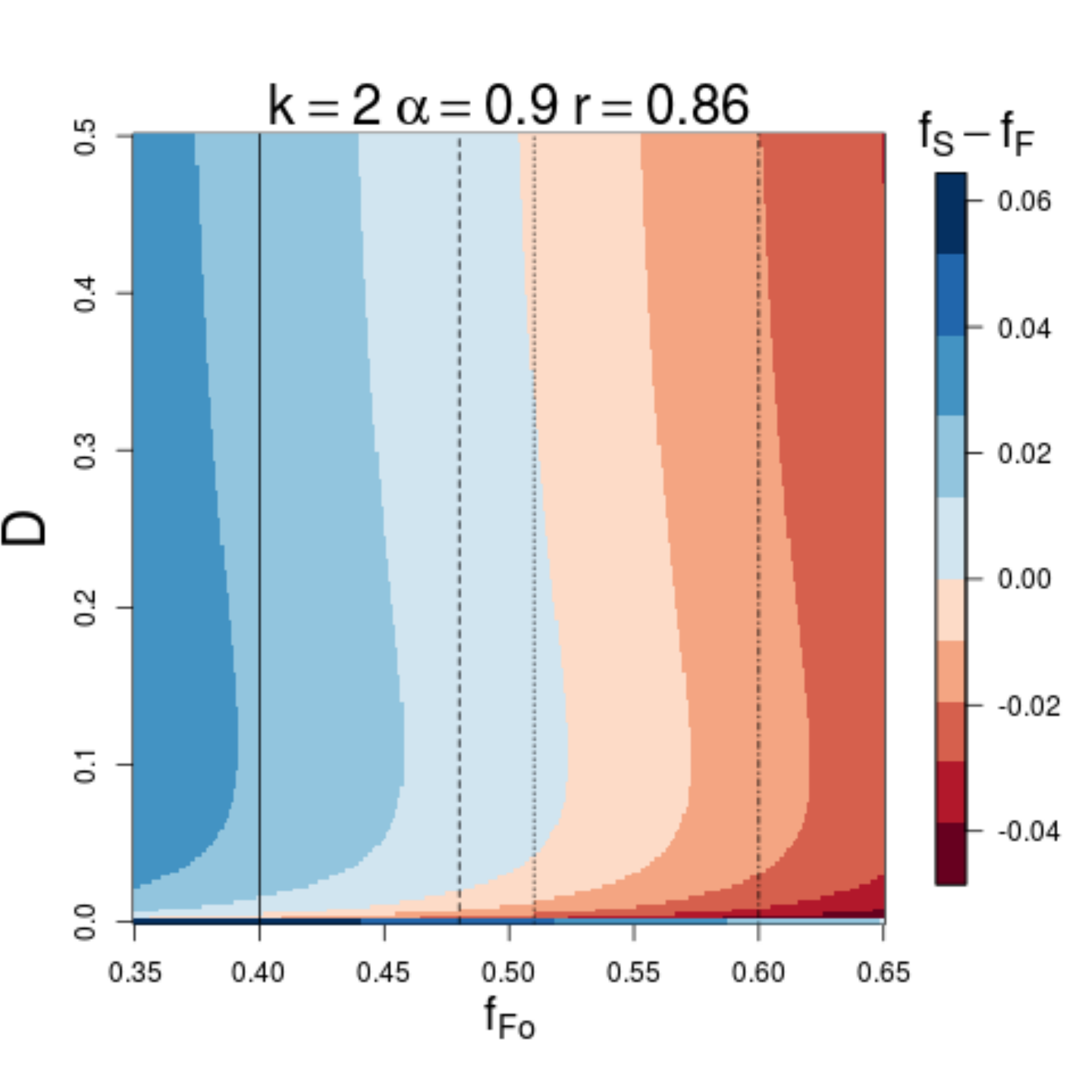}
\caption{Diagram of the relative difference $f_S-f_F$. The four vertical  straight lines in this diagram corresponds to the curves in the Fig.\ref{Fig:tot_vs_D_fao} with $f_{Fo}=\{0.4,0.48,0.51,0.6\}$.}
\label{Fig:diagram_vs_D_fao_0}
\end{figure}

Combining Figs.~\ref{Fig:tot_vs_D_fao}~and~\ref{Fig:diagram_vs_D_fao_0}, we understand the existence of an optimal value for each curve $f_{Fov} = {\rm constant}$; that is reminiscent of a competition mechanism between the multifold features of the model. For absent patch diffusion, $D=0$, the agents interact only inside one patch. For maximal diffusion, the individuals are able to interact with a much larger number of other individuals at the expense of weakening the links that necessarily establish the populations of the metapopulation.
Accordingly, there is an optimal value of the diffusion parameter, $D^{op}$, for each $f_{Fov}$ at which is achieved a balance between finding new individuals whilst preserving the robustness of the metapopulation. Notwithstanding, species $F$ is at its maximal situation, it does not mean it is the prevailing species as in dealing with a similar scenario species $S$ can end up in a situation for which the number of elements in the total population overcomes that $F$.

\begin{figure*}[!hbtp]
\centering 
\includegraphics[width=0.99\linewidth]{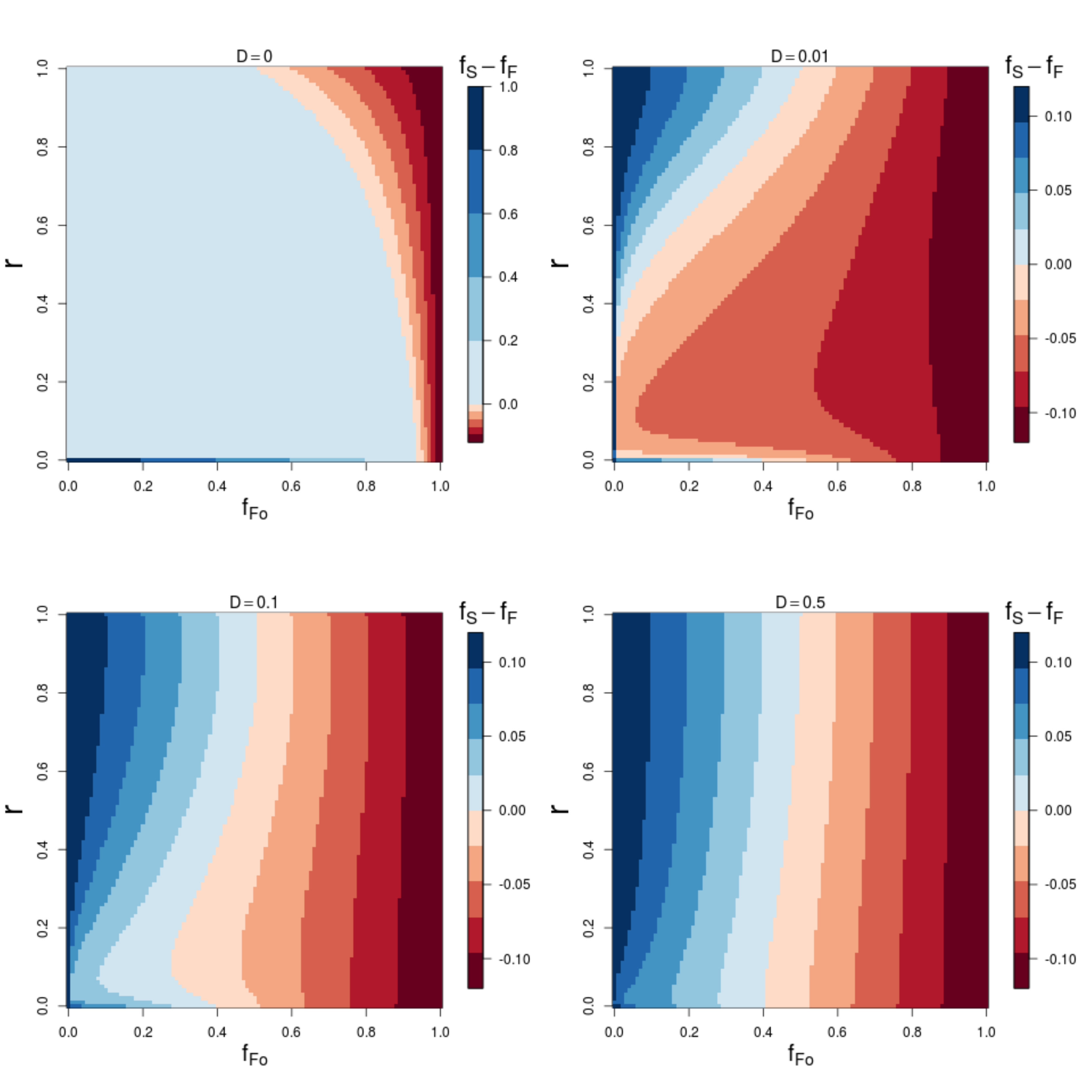}

\caption{Dependence of the relative difference $f_S-f_F$ in the steady state with $r$ versus $f_{Fo}$. Diagrams obtained from Eqs.~(\ref{Eq:Vu2})-(\ref{Eq:Bu2}) with $\alpha=0.88$, $k=2$.}
\label{Fig:r_vs_fao_D_1}
\end{figure*}

We have extended that analysis to other values of the  asymmetry parameter  $r$ keeping the value of $D$ constant in $D=\{0,0.01,0.1,0.5\}$ in Fig.~\ref{Fig:r_vs_fao_D_1}.  
When there is no diffusion in the system, $D=0$, $S$ is the dominant species, excepting for large values of $f_{Fo}$ and $r$. However, such scenarios change drastically for all $D>0$. Specifically, even for the case $D=0.01$ for which there is very little dispersal, it is already possible to change the outcome regarding the steady-state dominant species. For $r=0$, the dynamics of $S$ is naturally frozen because $\lambda_S=r\lambda_F=0$ and $\alpha_S=r\alpha_F=0$. The broad panorama from the present panels highlights that the establishment of the final majority depends on an intricate relation between competition, mobility and birth-death dynamics under spatial constraints.

\begin{figure}[!hbtp]
\centering 
\includegraphics[width=0.99\linewidth]{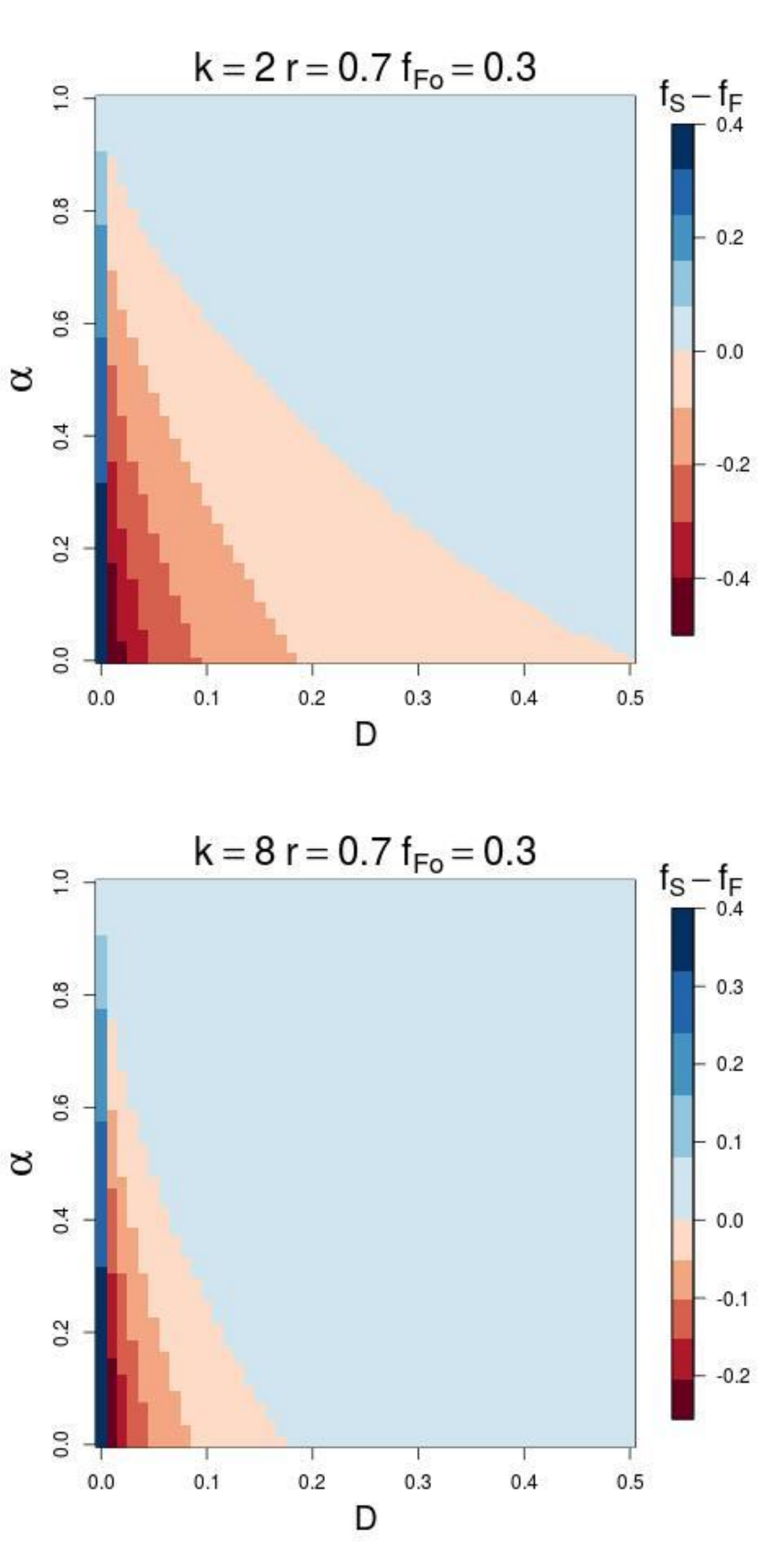}
\caption{Dominant species in the mortality vs dispersal diagram for strong ($k=2$) and weak ($k=8$) spatial constraints.}
\label{Fig:r_vs_fao_D_2}
\end{figure}

Finally, in the panels exhibited in Fig.~\ref{Fig:r_vs_fao_D_2} we analyze how the number of new patches that individuals can move impacts on the mortality-dispersal diagram for the competition between species F/S considering severe ($k=2$) and loose ($k=8$) spatial constraints. First, note that for  $D=0$ the slower species (S) becomes dominant
for any $\alpha$. Yet, this picture changes when we have mobility given by $D \neq 0$. For a given $\alpha$ not too high, $\alpha \lesssim 0.2$ , $S$ is dominant only if $D$ is high enough.  Increasing the number of new patches from $k=2$ to $k=8$, by reducing the spatial constraints, the individuals are naturally able to move more which leads to an enhancement of the advantage of being slow. To explain such results keep in mind that in the long-run the mobility spreads the absence of local correlations to the whole metapopulation and thus the results for high $D$ qualitatively approaches the results for $D=0$ as $k$ increases.

\section{\label{sec:fin-rem}Final remarks}

In this work, we have used diffusion to study the role played by mobility in quasi-neutral competition within a metapopulation context, which from a physical perspective can be understood as a coarse-grained approach to an ecological system. Considering quasi-neutral competition a metapopulation analysis is worthwhile since in being a population of populations, events that affect local populations, namely the possibility of moving with impact on mating and the structure of network correlations, can put the whole metapopulation structure.

Our theoretical results -- obtained from Monte Carlo simulations as well as numerical integration of multidimensional ODEs -- show that the multifold interplay between quasi-neutral competition between two species with different biological clocks, spatial constraints, and diffusion in metapopulation is remarkably complex. Nevertheless, it was possible to understand that large mobility between different patches --- which at first would benefit mating --- can have the same impact as no migration between patches. That being so, for given initial conditions, there is a set of parameters that optimize the population imbalance, which can be favorable to the slower species. In other words, depending on the level of mobility, being biologically slower can be actually an advantage.  The take-home message from our work is that the ecological majority-minority switching for quasi-neutral competition in metapopulations exhibits a nonmonotonic relation with the diffusion between patches. This phenomenon is highly counter-intuitive, but it could be further studied resorting to experimental setups within Synthetic Biology where bacteria can be programmed to exhibit new behaviour~\cite{smith2014programmed,ding2014synthetic,wang2016build,padilla2015synthetic}.
From a broader point of view, the present contribution adds an interesting and new building block to the list  of counter-intuitive  ecological dynamics
~\cite{shaw2015dispersal,lombardo2014nonmonotonic,khasin2012minimizing,korolev2015evolution,abbott2011dispersal,duncan2015dispersal,cheong2019paradoxical,doak2008understanding}. It is worthwhile to note that minority-majority inversions have also been observed in social systems~\cite{crokidakis2014first}.

Some points were not addressed in our work, like the impact of the presence of a topology or time-dependent rates in the results. In future works, it would be interesting to generalize our model to incorporate time-dependent dispersal rates as well as networks with more sophisticated topologies than those used in this manuscript.

\begin{acknowledgements}
We acknowledge the financial support from the Brazilian funding agencies CAPES, CNPq and FAPERJ.



\end{acknowledgements}


\noindent
\textbf{Conflict of interest} The authors declare that they have no conflict of interest.

\bibliographystyle{spphys}       
\bibliography{main.bib}

\end{document}